\documentclass[aps,prl,twocolumn,showpacs,10pt,superscriptaddress,preprintnumbers,nofootinbib]{revtex4-1}
\usepackage{epsfig,amssymb,amsmath,psfrag,epstopdf,color}
\usepackage{graphicx}
\usepackage{multirow}
\usepackage{ulem}

\newcommand{\TXT}{\textrm}

\newcommand{\be}{\begin{equation}}
\newcommand{\ee}{\end{equation}}
\newcommand{\bea}{\begin{eqnarray}}
\newcommand{\eea}{\end{eqnarray}}

\allowdisplaybreaks

\begin{document}

\title{Production and hadronic decays of Higgs bosons in heavy ion collisions}
\author{Edmond L. Berger}
\email{berger@anl.gov}
\affiliation{High Energy Physics Division, Argonne National Laboratory, Argonne, Illinois 60439, USA}
\author{Jun Gao}
\email{jung49@sjtu.edu.cn}
\affiliation{INPAC, Shanghai Key Laboratory for Particle Physics and Cosmology,
School of Physics and Astronomy, Shanghai Jiao Tong University, Shanghai 200240, China}
\author{Adil Jueid}
\email{adil.jueid@sjtu.edu.cn}
\affiliation{INPAC, Shanghai Key Laboratory for Particle Physics and Cosmology,
School of Physics and Astronomy, Shanghai Jiao Tong University, Shanghai 200240, China}
\author{Hao Zhang}
\email{zhanghao@ihep.ac.cn}
\affiliation{Theoretical Physics Division, Institute of High Energy Physics, Chinese Academy of Science, Beijing 100049, China}
\affiliation{School of Physics, University of Chinese Academy of Science, Beijing 100049, China}

\begin{abstract}
We examine Higgs boson production and decay in heavy-ion collisions at the LHC and future 
colliders. 
Owing to the long lifetime of the Higgs boson, its hadronic decays may 
experience little or no screening from the hot and dense quark-gluon plasma whereas 
jets from hard scattering processes and from decays of the electro-weak gauge
bosons and the top-quark suffer significant energy loss.
This distinction can lead to 
enhanced signal to background ratios in hadronic decay channels and thus, for example, 
provide alternative ways to probe the Yukawa 
coupling of the Higgs boson to the bottom quark and its lifetime.

\end{abstract}

\pacs{}
\maketitle

\noindent \textbf{Introduction.}
The successful operation of the CERN Large Hadron Collider (LHC) led to the discovery of the 
Higgs boson, the final piece of the standard 
model (SM)~\cite{Aad:2012tfa,Chatrchyan:2012xdj} of particle physics.
Precise measurements of the properties and couplings of the Higgs boson are now required for a 
refined understanding of the nature of electroweak symmetry breaking and for searches for new 
physics beyond the SM.
This pursuit has high priority at the ongoing LHC and future high-luminosity LHC (HL-LHC) projects, 
and it has motivated consideration of dedicated Higgs boson production
facilities~\cite{Gomez-Ceballos:2013zzn,Moortgat-Picka:2015yla,CEPC-SPPCStudyGroup:2015csa}.

These investigations focus on the properties of the Higgs boson in the 
vacuum.  However, most of the Higgs bosons in the early universe existed 
in a high-temperature and high-density environment \cite{Gamow:1946eb,Alpher:1948ve}. 
An understanding of the role of the Higgs boson in the early universe would be 
advanced through study of the Higgs boson not only in the vacuum, but also in an extreme
medium.   Heavy-ion collisions at the LHC, proposed to  study properties of the quark-gluon plasma 
(QGP), create an extreme environment with high temperature and density~\cite{Armesto:2015ioy}.  
They are well suited at the same time to study the behavior of the Higgs boson in a hot dense 
environment.

The expansion and cooldown of the QGP at the LHC is predicted to have a typical
time scale of about $10\,{\rm fm/c}$~\cite{Connors:2017ptx,Pasechnik:2016wkt,Shen:2012vn}.
Although longer than the lifetime of the electro-weak (EW)
gauge bosons and the top-quark,  this time scale is shorter than the lifetime of the Higgs boson
(which is $\sim 47 \,{\rm fm/c}$).  The consequences include 
\begin{itemize}
\item {Particles from Higgs decay, which do not travel in the QGP, will carry 
information on the Higgs boson.}
\item {Because the strong backgrounds are reshaped by the QGP medium while the 
signal is nearly unchanged, the phenomenology of Higgs boson hadronic decay is different 
from $pp$ collisions.}
\item {A check of the first two consequences serves as a natural probe of the Higgs 
boson lifetime.}
\end{itemize}

In this Letter we study the production and decays
of the Higgs boson in heavy-ion collisions. 
We point out the main differences with the proton-proton case.
Jets produced from hadronic decays of the Higgs boson are not
affected much by the QGP since the decay happens at a much later stage.
Meanwhile, jets produced from hard QCD scattering and decays of EW
gauge bosons and the top-quark experience energy loss through interaction 
with the medium~\cite{Apolinario:2017sob}, known as jet quenching, an 
established phenomenon in heavy-ion collisions at the Brookhaven RHIC 
facility and the LHC~\cite{Qin:2015srf}.
These different responses lead to suppression of the SM backgrounds to 
hadronic decays of the Higgs boson and also to distinct kinematic configurations
of the signal and backgrounds, resulting in an enhanced ratio of the signal over
the background when compared to $pp$ collisions.  We explore different models of 
jet quenching to provide quantitative estimates for the case of $ZH$ associated 
production with Higgs decay $H \rightarrow b \bar{b}$. A different 
perspective on Higgs boson physics in heavy ion collisions is proposed in Refs.
\cite{dEnterria:2017jyt,dEnterria:2017iew}.

\noindent \textbf{Higgs boson production.}
The cross section for Higgs boson production in collisions of two heavy nuclei with charge $Z$ and 
atomic number $A$ is 
\begin{align}
  \sigma(AA\to H+X)&=A^2 c(f)\sum_{a,b} \int \TXT{d} x_a \TXT{d} x_b \nonumber \\
  &\hspace{-1in} \times f_{a/A}(x_a,\mu_F^2) f_{b/A}(x_b,\mu_F^2)  
 \hat{\sigma}(ab\to H+X).   
\end{align}
Here $f_{i/A}(x_i,\mu_F^2)$ is the effective nuclear parton distribution
function (PDF) of parton $i$ carrying
momentum fraction $x_i$ of the nucleon at a factorization scale $\mu_F$; 
$\hat{\sigma}$ is the partonic cross section; and 
$A^2c(f)$ is the number of nucleon collisions for the
centrality range $f$, for which $c(f)$ can be obtained by
integrating the overlap function of the two nuclei over
the corresponding range of impact parameters~\cite{Loizides:2017ack}.
For the centrality range 0-10\% in this study,
$c(f)$ is calculated to be 42\% with the Glauber Monte Carlo
model~\cite{Loizides:2017ack} 
for PbPb collisions at $\sqrt {s_{\text{NN}}}=$ 5.5 TeV.
In Table~\ref{tab:hxsec} we show cross sections for Higgs
boson production in different channels for PbPb collisions at
the LHC, HE-LHC, and FCC-$hh$~\cite{Mangano:2017tke} or SPPC~\cite{CEPC-SPPCStudyGroup:2015csa}, with $\sqrt {s_{\text{NN}}}$
=5.5, 11, and 39.4 TeV respectively.
We calculate the partonic cross sections with \textsc{MCFM}~
\cite{Campbell:2016jau,Boughezal:2016wmq} to next-to-leading
order in QCD for vector boson fusion (VBF) and next-to-next-to-leading
order (NNLO) for gluon fusion (GF) and for associated production. 
The cross sections for production in gluon fusion agree
well with those shown in
Ref.~\cite{dEnterria:2017jyt} apart from differences due to scale choices.
The centrality factors are similar for the three energies and are
not applied in Table~\ref{tab:hxsec}. 
For comparison, cross sections for production in $pp$ collisions are also listed in
Table~\ref{tab:hxsec}.
\begin{table}[!h]
\caption{
Cross sections for Higgs boson production from 
different processes in PbPb collisions and proton-proton
collisions at $\sqrt {s_{\text{NN}}}=$ 5.5, 11, and 39.4 TeV, respectively.
The nCTEQ15 PDFs~\cite{Kovarik:2015cma} and CT14 
PDFs~\cite{Dulat:2015mca}
are used for the PbPb and $pp$-collisions, respectively.
}
\centering
\vspace{2ex}
\begin{tabular}{c|ccc}
\hline\hline
\multirow{2}{*}{process}&
\multicolumn{3}{c}{PbPb($pp$) in nb(pb)}\\
& $\quad$ 5.5 TeV $\quad$ & $\quad$ 11 TeV $\quad$ &
 $\quad$ 39.4 TeV $\quad$ \\
\hline
GF &   ${480}({10.2})$&${1556}({35.2})$&${9580}({235})$\\
\hline
VBF & ${15.3}({0.316})$ & ${65.6}({1.40})$ & ${421}({10.02})$ \\
\hline
$ZH$ &  ${10.2}({0.230})$&${28.1}({0.687})$&${147}({3.97})$\\
\hline
$W^+H$ & ${8.38}({0.162})$&${21.8}({0.716})$&${94.2}({3.19})$\\
\hline
$W^-H$ & ${9.22}({0.143})$&${23.4}({0.435})$&${99.5}({2.34})$\\
\hline\hline
\end{tabular}
\label{tab:hxsec}
\end{table}

We focus on decays of the Higgs boson to bottom
quarks for which the associated production with a $Z$ boson and
its subsequent leptonic decay gives the strongest sensitivity
\cite{Aaboud:2017xsd,Sirunyan:2017elk},
albeit with a relatively small cross section.
The dominant backgrounds in this case are $Z$ plus bottom-quark pair 
production and top-quark pair production with leptonic decays.
Bottom quarks from decays of the Higgs boson form two energetic jets that
can be detected with various $b$-tagging algorithms~\cite{Chatrchyan:2013exa}. 
On the other hand, in the environment of heavy-ion collision, $b$-jets
from the backgrounds will lose energy from interactions in the  QGP~\cite{Qin:2015srf}.       
Owing to the dead-cone effect of QCD radiation~\cite{dEnterria:2010ubj}, 
it has been argued that a primary $b$-quark will lose less energy than light quarks 
when traversing QGP,  but 
experimental measurements have shown similar level of nuclear
suppression for inclusive jets and $b$-jets, and similar distortion of transverse
momentum balance~\cite{Chatrchyan:2013exa,Sirunyan:2018jju} 
of dijets from jet quenching.
The fraction of energy lost from a primary $b$-quark jet is thus believed to be 
comparable to that from a light quark, at least for jets with high transverse momentum.
There are also theoretical studies supporting the similarity
of quenching of jets initiated by $b$-quark and light
quarks~\cite{Huang:2013vaa,Djordjevic:2016vfo,Senzel:2016qau}.

\noindent \textbf{Jet Quenching Models.}
We base our quantitative estimates on simplified 
phenomenological models of jet quenching since a full Monte Carlo generator with jet 
quenching is not available for the processes of interest \footnote{Such generators exist for 
QCD jets production, prompt photon
production and electroweak boson plus a single jet
production~\cite{Arneodo:1996ra,Lokhtin:2008xi,Zapp:2013vla}.}.
Differences among the three models provide a measure of the uncertainties in our results.  
%
\begin{figure}[!h]
  \begin{center}
  \includegraphics[width=0.5\textwidth]{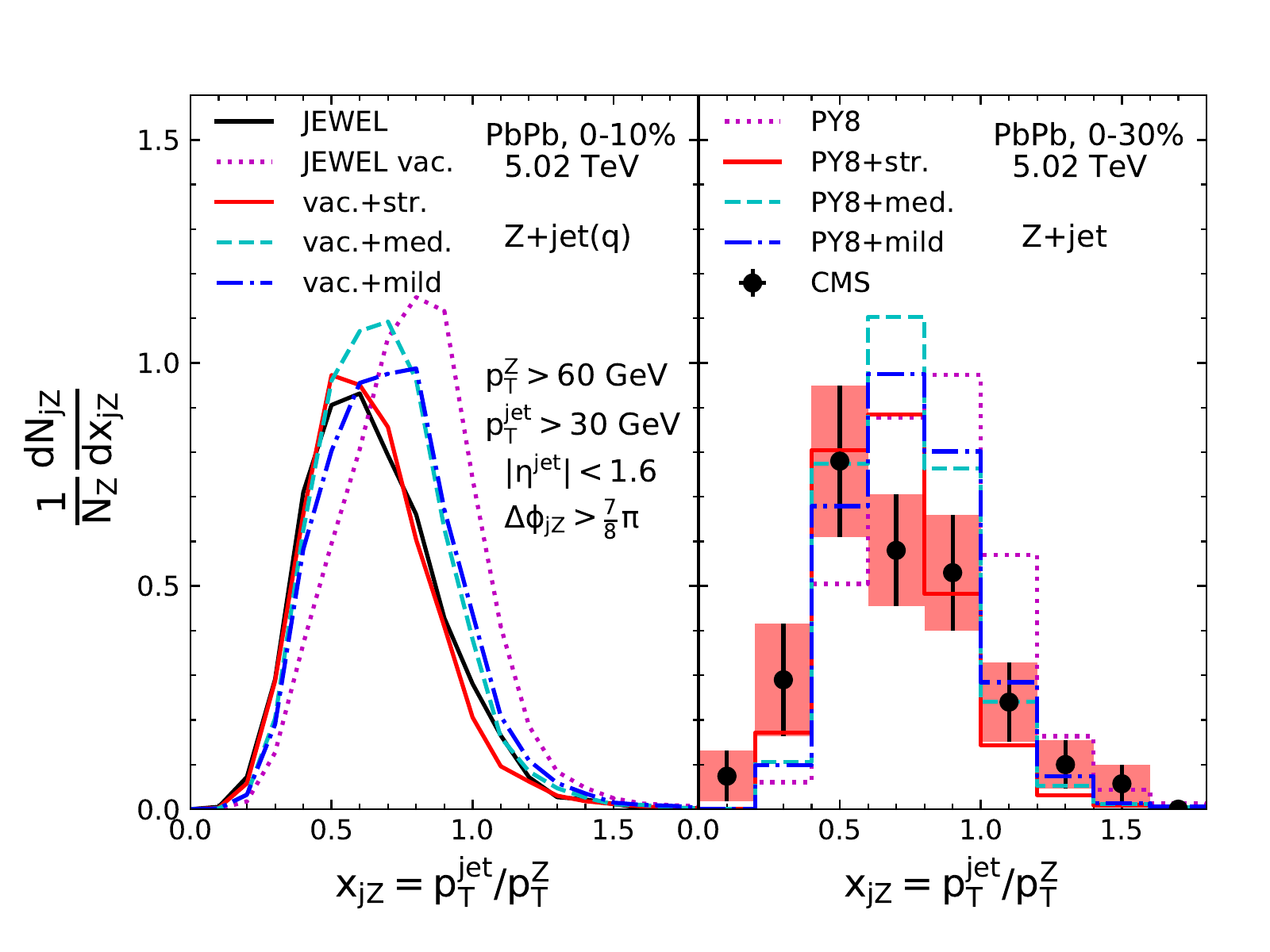}
  \end{center}
  \vspace{-2ex}
  \caption{\label{fig:fig1a}
  The impact of different models on jet observables is shown taking as an example, 
  production of a $Z$ boson plus a single jet. 
  Distribution of the ratio of transverse momenta in $Z+jet$ production
  in PbPb collisions. 
  Left: comparison of predictions from \textsc{Jewel}2.0 and the folded results
  with various models, for centrality class 0-10\% and only quark final
  states included;
  right: comparison of the folded results with CMS measurement for
  centrality class 0-30\%.   
  }
\end{figure}
The average loss of transverse momentum for a jet traversing the QGP compared to the 
vacuum is parametrized with a convenient form
\begin{equation}\label{eq:quenching}
\langle\delta  p_{\text{T}} \rangle=a p_{\text{T}}+b\ln(p_{\text{T}}/{\rm GeV})+c.  
\end{equation}  
The parameters depend on the center of mass energy, the collision
centrality, and also the jet reconstruction scheme.
In the following we use the anti-$k_{\text{T}}$~\cite{Cacciari:2008gp} algorithm with $R=0.3$.
The choice of small jet cone size is typical for heavy-ion collision in order
to minimize effects of fluctuations due to underlying events.
We choose three representative models for quark jets in PbPb collisions
with a centrality class of $0-10$\%, i.e., with strong quenching $a=0$,
$b=2$ GeV, $c=12$ GeV, medium quenching $a=0.15$, $b=c=0$, and mild
quenching $a=b=0$, $c=10$ GeV.
These choices correspond to a loss of transverse momentum of 21, 15, and 10 GeV
respectively, for a jet with $p_{\text{T}}=100$ GeV in vacuum.
The model with medium quenching was used previously in a study of top-quark pair
production in heavy-ion collisions~\cite{Apolinario:2017sob} except there the
scaling was applied on individual constituent particles.           
In addition we impose Gaussian smearing on the energy loss to mimic
the fluctuations in jet quenching with width set to half of the average energy lost.
The jet energy resolution is parametrized as
\begin{equation}\label{eq:resolution}
\sigma(p_{\text{T}})=\sqrt{C^2+\frac{S^2}{p_{\text{T}}}+\frac{N^2}{p_{\text{T}}^2}}.
\end{equation}
Representative values of the $C$, $S$, and $N$ parameters from
CMS for different centrality classes in PbPb collisions can be found in
\cite{Sirunyan:2017qhf} and are used in our calculations.
In PbPb collisions, subtraction of underlying events is performed
and contributes to the $N$ term in jet energy resolution and also a deterioration
of the $S$ term. 

The transverse momentum imbalance in $Z$ boson plus jet production was measured 
recently by the CMS collaboration in PbPb collisions 
at $\sqrt{s_{\text{NN}}}=5.02$ TeV as a hard probe of jet quenching~\cite{Sirunyan:2017jic}.
Following the analysis in ~\cite{Sirunyan:2017jic}, we plot in 
Fig.~\ref{fig:fig1a}  
distributions of the ratio of the transverse momenta 
$x_{jZ}=p_{\text{T}}^{jet}/p_{\text{T}}^{Z}$ normalized to the rate of
inclusive $Z$ boson production, where $p_{\text{T}}^{jet}$ is
the transverse momentum of the leading jet.
In the plot on the left side of Fig.~\ref{fig:fig1a} we show predictions 
from the Monte Carlo program \textsc{Jewel} 2.0.0 \cite{Zapp:2013vla} for the
centrality class 0-10\%.
A prediction without jet quenching (vacuum) is also shown, obtained from 
 \textsc{Pythia 6.4} \cite{Sjostrand:2006za} incorporated in \textsc{Jewel} 2.0.0.
We use only the hard matrix elements for quark final states since we are
interested in quenching of jets initiated by quarks.
The initial temperature of 
the QGP is set to 590 MeV~\cite{KunnawalkamElayavalli:2016ttl}.  
A shift to lower values is seen in the distribution as quenching is increased, 
as well as a reduction of the event rate. 
For comparison with the \textsc{Jewel} prediction, we also show predictions obtained by applying 
our quenching models to the vacuum calculation on a event-by-event basis.  
The folded result with strong quenching is in good agreement with the \textsc{Jewel} result.
In the plot on the right of Fig.~\ref{fig:fig1a} we compare our folded results with the CMS data measured 
for centrality class 0-30\%~\cite{Sirunyan:2017jic}.
The baseline vacuum prediction is from \textsc{Pythia} 8~\cite{Sjostrand:2014zea} with
both gluon and quark final states included; the latter contributes more than 80\% of the total production rate.
The CMS data disfavor the vacuum prediction.  The three simplified quenching models are consistent
with current data.
%

\noindent \textbf{Signal and backgrounds.}
We consider the signal process PbPb$\to ZH \to\ell^+\ell^-b\bar b$, in the 0-10\%
centrality class, with $\ell=e,\mu$ for which the QCD backgrounds are highly
suppressed. 
We simulate the signal and backgrounds at leading order using
\textsc{sherpa} 2.2.4~\cite{Gleisberg:2008ta} including
parton showering and hadronization, and with nCTEQ15 PDFs\cite{Kovarik:2015cma}.
The dominant SM backgrounds are $Zb\bar b$ production and $t \bar{t}$
production with leptonic decays of top quarks.
Other SM backgrounds including those from production of $Z$ plus light flavors are
significantly smaller and are ignored.
We normalize the total cross sections of the signal to the NNLO values in 
Table~\ref{tab:hxsec}, and of the $t\bar t$ background to the NNLO predictions
with resummed corrections from \textsc{Top++2.0}~\cite{Czakon:2011xx,Czakon:2013goa},
times the relevant centrality factors. 
The Monte Carlo events are passed to \textsc{Rivet}~\cite{Buckley:2010ar} for
analysis with an anti-$k_{\text{T}}$ jet algorithm as implemented in \textsc{Fastjet}~\cite{Cacciari:2011ma} and a distance parameter of 0.3. 
Jet quenching and jet energy resolution are applied according
to Eqs.(~\ref{eq:quenching}) and (~\ref{eq:resolution}).  
We use pre-selection cuts similar to those in the
CMS heavy-ion analysis~\cite{Sirunyan:2017jic},
\begin{eqnarray}
&&p_{\text{T}}^\ell>15\,{\text{GeV}},\quad|\eta^\ell|<2.5,
\quad\Delta R_{\ell\ell}>0.2,\nonumber\\
&&p_{\text{T}}^{j}>30\,{\text{GeV}},\quad |\eta^j|<1.6,\quad\Delta R_{j\ell}>0.3.
\end{eqnarray}
We select events in the following signal-like region 
\begin{itemize}
\item A pair of same-flavor opposite-sign charged
leptons with invariant mass $|m_{\ell\ell}-m_Z|<10$ GeV;
\item Exactly two jets, both $b$-tagged,  
with separation $\Delta R_{bb}<2.0$;
\item The transverse momentum of the reconstructed
vector boson $p_{\text{T}}^Z\equiv p_{\text{T}}^{\ell\ell}>100$ GeV.
\end{itemize}
We assume a $b$-tagging efficiency of 80\%, better than that 
achieved in the CMS analysis~\cite{Chatrchyan:2013exa}, 
but expected in future runs.
The requirement of large $p_{\text{T}}^Z$ can suppress the
$t\bar t$ background efficiently.

%
\begin{figure}[!h]
  \begin{center}
  \includegraphics[width=0.48\textwidth]{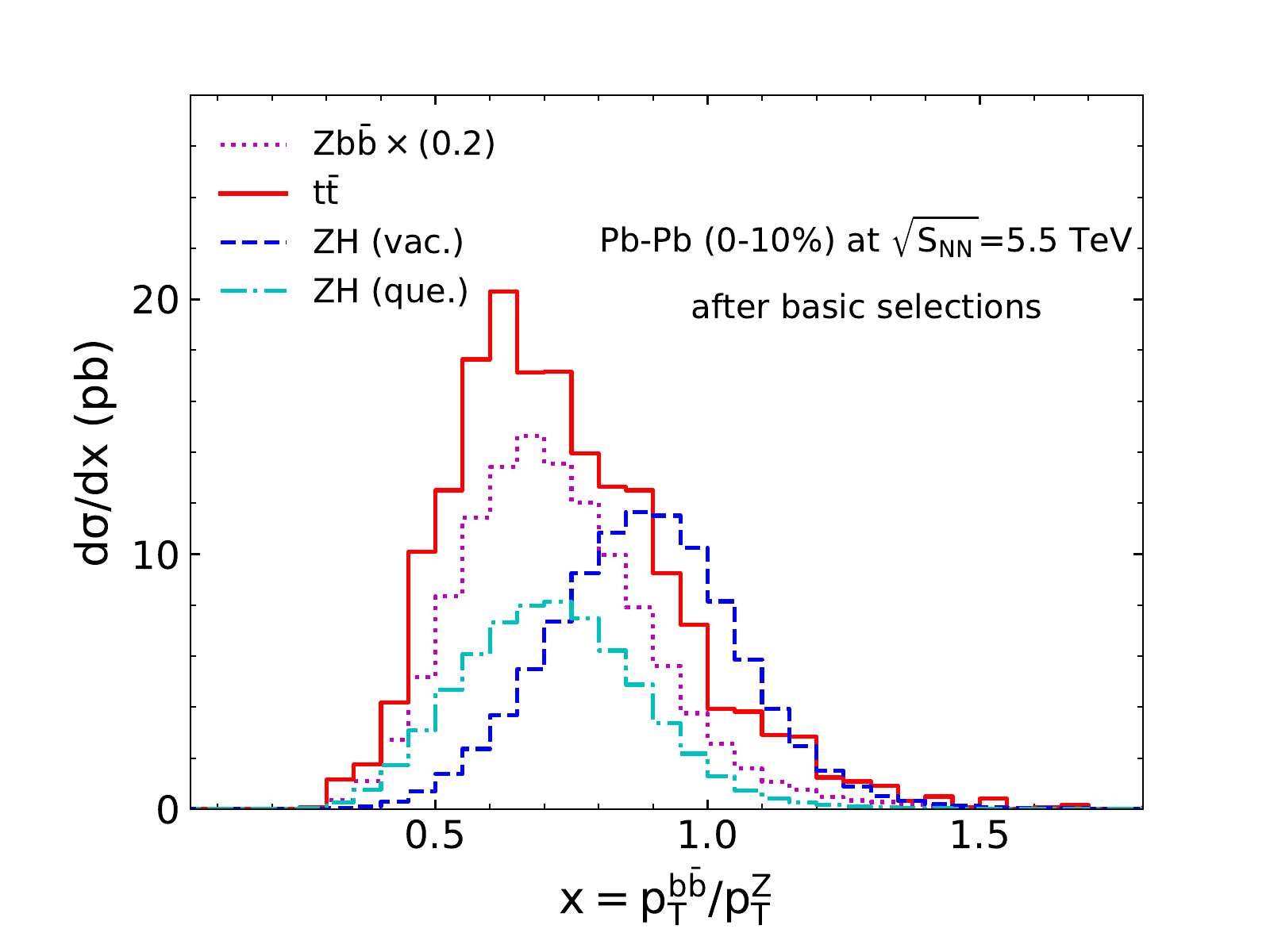}
  \end{center}
  \vspace{-2ex}
  \caption{\label{fig:fig2}
  Distributions of the ratio of the transverse momenta of the pair of $b$-jets and the
  $Z$ boson for PbPb collision with $\sqrt{s_{\text{NN}}}=5.5$ TeV and
  centrality class 0-10\%, after basic selections.
  For the nominal case both backgrounds are strongly quenched while the signal 
  in unquenched.
  The distribution for a quenched signal is also shown as a comparison. 
  The $Zb\bar b$ result has been multiplied by 0.2.
  }
\end{figure}

The analysis so far follows Ref.~\cite{Aaboud:2017xsd}.
As mentioned earlier, different quenching properties of the signal and
backgrounds lead to further separation in certain variables.
Separation is illustrated in Fig.~\ref{fig:fig2} for the ratio $x=p_{\text{T}}^{b\bar b}/p_{\text{T}}^Z$
of the transverse momenta of the reconstructed $b\bar b$ pair and the $Z$ boson.
We apply the strong quenching model on the two backgrounds and the signal is
vacuum-like.
The backgrounds tend to peak in the region of smaller $x$  since both of the
$b$-jets lose a fraction of their energies.
In Fig.~\ref{fig:fig2}, we also show the result for the extreme case in which the 
$b$-jets in the signal process are also strongly quenched.
In this case, besides the shift of the peak, the signal normalization is also 
reduced since more $b$-jets fall below the $p_{\text{T}}$ threshold.
Not shown here, we find that the transverse momentum of the (sub)leading-jet shows 
similar separation power.     

\begin{figure}[!h]
  \begin{center}
  \includegraphics[width=0.48\textwidth]{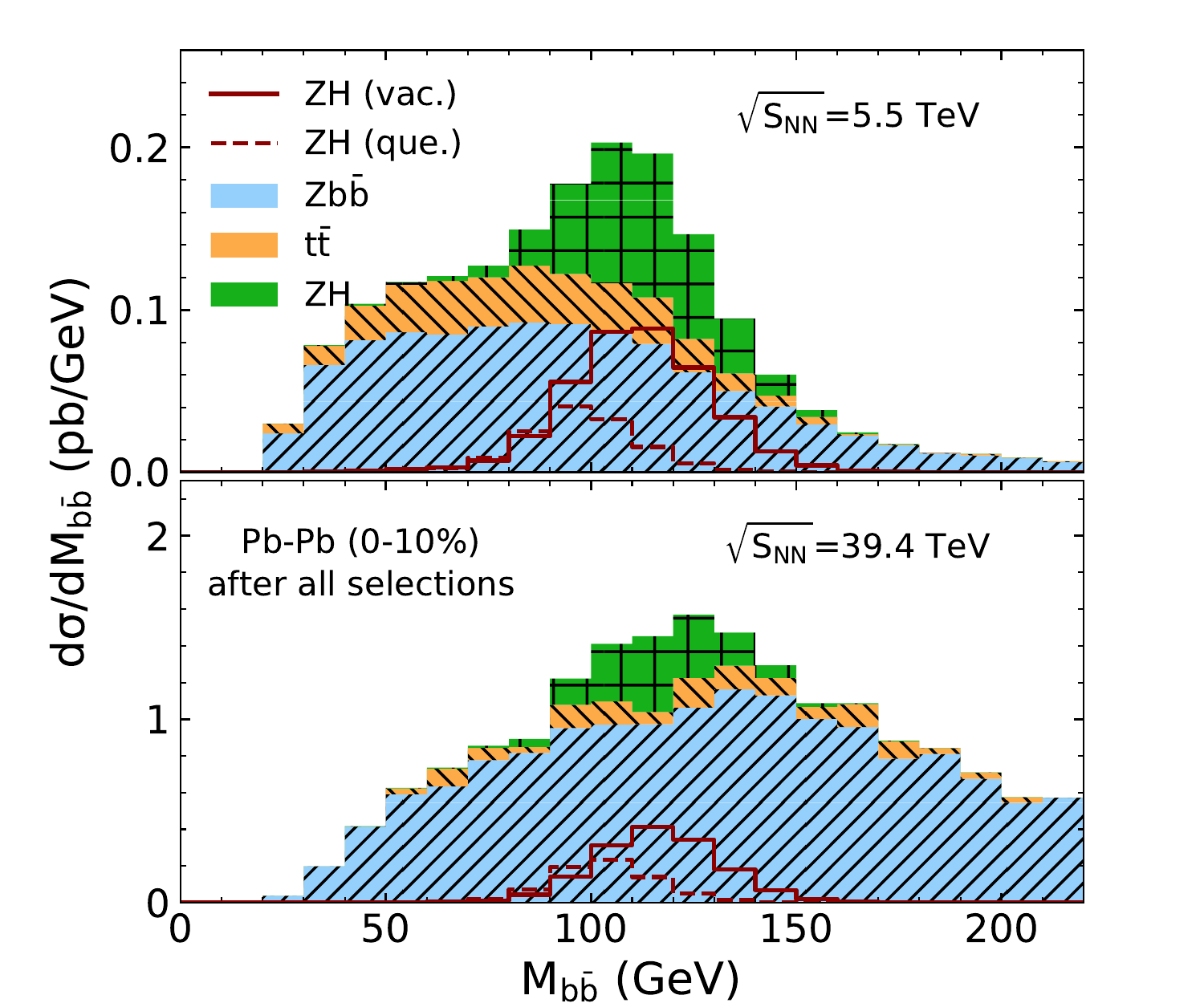}
  \end{center}
  \vspace{-2ex}
  \caption{\label{fig:fig3}
  Distributions of the invariant mass of the pair of $b$-jets after all selections,
  similar to Fig.~\ref{fig:fig2}.
  }
\end{figure}

To establish the discovery potential of the signal we demand 
events with $x>0.75$, and $p_{\text{T}}>60$ GeV for the leading-jet at
LHC and HE-LHC and for the subleading-jet at FCC-$hh$.  
The invariant-mass distribution of the two $b$-jets $M_{b\bar b}$ is 
shown in Fig.~\ref{fig:fig3} after all selections.
The dominant background is $Zb\bar b$, and the signal exhibits
a clear peak near the Higgs boson mass.
The large width of the signal reflects the effects of jet energy smearing.
In Fig.~\ref{fig:fig3} we also display the signal distribution for the case 
of strong quenching.  It shows a much weaker peak at lower mass.  
Comparison of FCC-$hh$ to LHC shows that the background to signal ratio increases for $Zb\bar b$
owing to the higher energy and decreases for $t\bar t$ as a result of the cut on subleading-jet.   
%
\begin{figure}[!h]
  \begin{center}
  \includegraphics[width=0.48\textwidth]{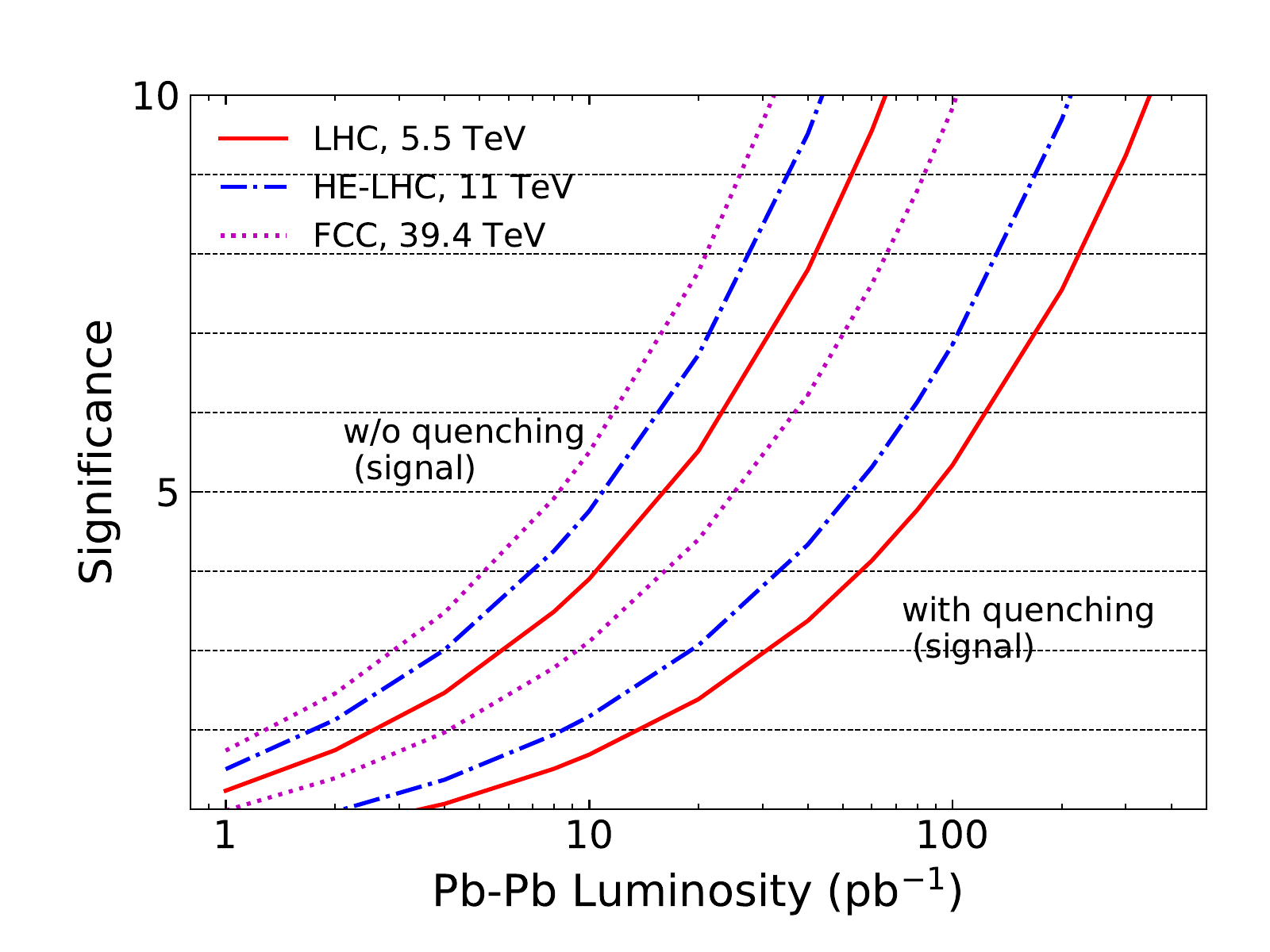}
  \end{center}
  \vspace{-2ex}
  \caption{\label{fig:fig4}
  Expected significance of the Higgs boson signal as a function of
  ion luminosity for PbPb collisions at LHC, HE-LHC, and FCC-$hh$.
  Results for the case of a quenched signal are also shown for comparison.
  }
\end{figure}

We use the log-likelihood ratio $q_0$~\cite{Cowan:2010js} as a test-statistic to calculate
the expected significance of the signal based on the $M_{b\bar b}$ distribution,
as a function of the integrated luminosity of the collision program.
The results are shown in Fig.~\ref{fig:fig4} and in Table~\ref{tab:signif}.
For the LHC, a $5(3)\sigma$ discovery(evidence) requires a total ion luminosity
of about 16(5.9) $\rm pb^{-1}$ in PbPb collisions, larger than the projected
LHC luminosity~\cite{Jowett:2008hb}.
The numbers are 8.0(2.9) $\rm pb^{-1}$ for PbPb collisions at FCC-$hh$.
The significance if the signal is also quenched
are much lower than the nominal case shown in Fig.~\ref{fig:fig4}.       
The results for alternative quenching models and for no quenching of the 
backgrounds are summarized in Table~\ref{tab:signif}.
The improvement in signal-background discrimination from jet quenching is clear.
We expect the sensitivity can be further improved for example by using multi-variate
analysis and by including $Z$ decays into neutrinos and $WH$ production
as demonstrated in Ref.~\cite{Aaboud:2017xsd}.
Taken together they may bring down the needed luminosity by a factor of two.
Nevertheless, with a much lower luminosity one can manage to study Higgs boson
production in the diphoton channel~\cite{dEnterria:2017jyt,dEnterria:2017iew},
including its interaction with the medium~\cite{,dEnterria:2018new}.    

%
\begin{table}[!h]
\caption{
Ion luminosity required to reach 5$\sigma$ significance for 
the signal for different models of jet quenching and collision
energies. 
Numbers in parenthesis correspond to a 3$\sigma$ evidence. 
}
\centering
\vspace{2ex}
\begin{tabular}{c|cccc}
\hline\hline
 $\quad$lumi.($\rm pb^{-1}$) $\quad$&  \, strong \, & 
 \, medium \, & \, mild\, &  \, vacuum \,  \\
\hline
LHC &   $16(5.9)$ &$27(9.8)$&$26(9.3)$ & $48(17)$\\
\hline
HE-LHC & $11(4.0)$ & $20(7.2)$ & $20(7.2)$ & $34(12)$\\
\hline
FCC-$hh$ &  $8.0(2.9)$& $13(4.7)$ & $14(5.0)$& $22(8.0)$\\
\hline\hline
\end{tabular}
\label{tab:signif}
\end{table}
 
\noindent \textbf{Summary.}
The long lifetime of the Higgs boson relative to the typical time scale of the QGP makes 
it plausible that the strong decay products of Higgs bosons produced in heavy ion collisions 
escape the QGP medium unaffected.
On the other hand, QCD backgrounds will be attenuated 
by jet quenching.
These features open the possibility of enhanced ratios of signal to backgrounds.  
We demonstrated these ideas with the specific example of associated $Z H$ production in PbPb 
collisions at various colliders using simplified models of jet quenching.
The integrated luminosities 
needed for an observation of the signal are $\sim 10\,{\rm pb}^{-1}$.
It will be interesting to investigate the 
potential of other production channels of the Higgs boson with larger cross
sections~\cite{dEnterria:2017jyt,dEnterria:2017iew,Berger:2018new},
and also the possibility of using information on jet
shapes~\cite{Chatrchyan:2013kwa,Chien:2015hda,Apolinario:2017qay,Li:2017wwc}
expected to be different for quenched and unquenched jets.

\begin{acknowledgments}
JG thanks Lie-Wen Chen for useful discussions.
ELB's work at Argonne is supported in part by the U.S. Department of Energy under 
Contract No. DE-AC02-06CH11357.
The work of
JG and AJ is supported by the National Natural Science Foundation of China under
Contract No. 11875189 and No. 11835005.
The work of HZ is supported by Institute of High Energy Physics, Chinese Academy of
Science,  under Contract No. Y6515580U1
and Innovation Grant Contract No. Y4545171Y2.
\end{acknowledgments}

\bibliographystyle{apsrev}
\bibliography{higgsqgp}
\end{document}